\documentclass{elsart}
\usepackage{epsfig}
\begin{document}
\title{Fluctuations, correlations and the nonextensivity}
\author{G. Wilk$^{a}$, Z.W\l odarczyk$^{b,c}$}
\address{$^{a}$The Andrzej So\l tan Institute for Nuclear Studies,
                Ho\.za 69; 00-689 Warsaw, Poland\\ e-mail: wilk@fuw.edu.pl\\
         $^{b}$Institute of Physics, \'Swi\c{e}tokrzyska Academy,
               \'Swi\c{e}tokrzyska 15; 25-406 Kielce, Poland;\\
         $^{c}$ University of Arts and Sciences (WSU), Weso\l a 52,
                25-353 Kielce, Poland\\
                e-mail: wlod@pu.kielce.pl\\
 \today}
{\scriptsize Abstract: Examples of joint probability distributions are
studied in terms of Tsallis' nonextensive statistics both for correlated
and uncorrelated variables, in particular it is explicitely shown how
correlations in the system can make Tsallis entropy additive and that the
effective nonextensivity parameter $q_N$ decreases towards unity when the
number of variables $N$ increases. We demonstrate that Tsallis
distribution of energies of particles in a system leads in natural way to
the Negative Binomial multiplicity distribution in this system.


}

\section{Introduction}

It is very well known fact that whenever in a single variable exponential
distribution
\begin{equation}
f(x) = \frac{1}{\lambda}\cdot \exp\left( - \frac{x}{\lambda}\right)
\label{eq:f(x)}
\end{equation}
parameter $\lambda$ fluctuates according to gamma distribution, i.e.,
$\lambda \rightarrow \lambda' = \lambda' (\varepsilon) =
\frac{\lambda}{\varepsilon}$, where $\varepsilon$ is distributed
according to
\begin{equation}
g(\varepsilon) = \frac{1}{(q-1)\Gamma \left(\frac{1}{q-1}-1\right)}
\left[ \frac{\varepsilon}{q-1} \right]^{-2 + \frac{1}{q-1}} \cdot \exp
\left( - \frac{\varepsilon }{q-1} \right),
 \label{eq:lambdaprim}
\end{equation}
one obtains as result the following power-like distribution \cite{BC,WW}:
\begin{equation}
h(x) = \int^{\infty}_0 \!\!d\varepsilon\, g(\varepsilon)\left(
\frac{\varepsilon}{\lambda}\right) \exp \left[ -
\frac{\varepsilon}{\lambda}\cdot x \right] = C_1 \left[1 -
(1-q)\frac{x}{\lambda}\right]^{\frac{1}{1-q}}; ~ C_1=\frac{2-q}{\lambda}
, \label{eq:h(x)}
\end{equation}
called Tsallis distribution and characterized by parameter $q$
\cite{T,EuroPhys} ($q\in (1,2)$, for $q\rightarrow 1$ eq. (\ref{eq:h(x)})
becomes the usual exponential distribution given by eq. (\ref{eq:f(x)})).
Actually fluctuations can be also described by more general distributions
(which induce wide spectrum of the so called {\it superstatistics}
\cite{BC}), in all cases parameter $q$ reflects the amount of
fluctuations and is connected with their measure given by $\omega =
Var(\varepsilon)/\langle \varepsilon\rangle^2 $. In the case of eqs.
(\ref{eq:lambdaprim},\ref{eq:h(x)}) one has $\langle \varepsilon\rangle =
2 - q$, $Var(\varepsilon)=(q-1)\langle \varepsilon\rangle$ and
\begin{equation} \omega =
\frac{\langle \varepsilon^2\rangle}{\langle \varepsilon\rangle^2} - 1 =
\frac{q - 1}{2-q}\qquad {\rm or}\qquad q = 1
+\frac{\omega}{1+\omega}.\label{eq:defq}
\end{equation}
This is result for the so called type B superstatistics \cite{BC}. In
type A superstatistics,  not accounting for $\lambda$-dependent
normalization \cite{WW}, one gets $q=1+\omega$. In this case large
fluctuations were corresponding to large $q$, whereas eq. (\ref{eq:defq})
describes all fluctuations using only limited range of parameter $q$,
ranging from $\omega =0$ for $q=1$, up to $\omega \rightarrow \infty $
for $q$ reaching its maximal allowed value of $q \rightarrow 2 $ (notice
that for small values of $\omega$ both approach lead practically to the
same result: $q\simeq 1 +\omega$).

Such distributions are widely used to characterize systems with
stochastic pro\-cesses and are often associated with the existence of
long-range correlations, with memory effects and with nontrivial
(multi)fractal phase space structure \cite{T,EuroPhys}. Especially
important factor in getting such distributions are all possible intrinsic
fluctuations which exist in the system under consideration \cite{WW}
\footnote{In fact in recent work \cite{BGGM} it was argued that Tsallis
distribution does not imply dynamics with correlated signals but rather
with signals occurring in nonstationary intervals of time.}. In this
paper we shall study examples of formal interrelations between
fluctuations, correlations and nonextensivity, which can be of interest
in physical applications, especially in the field of high energy
multiparticle production processes (cf., \cite{WW,T,EuroPhys} and
references therein for other dynamical motivations). Notion of
nonextensivity reflects the fact that Tsallis distribution
(\ref{eq:h(x)}) has also its origin in the so called nonextensive
statistical mechanics (or information theory) based on Tsallis entropy,
which depends on parameter $q$ (becoming for $q\rightarrow 1$ the usual
Boltzmann-Gibbs-Shannon entropy). It is normally nonextensive by amount
proportional to $q-1$, therefore $q$ is named nonextensivity parameter
\cite{T,EuroPhys}. Anticipating some applications we shall in what
follows have in mind distributions of particles or their energies rather
than some unspecified variables. Our goal is limited, the more general
(but also very specialized) discussions on the role of correlations in
obtaining Tsallis distributions can be found in \cite{Kodama}, whereas in
\cite{CT-NEXT2005,CT,TG-MS} one can find further discussions dealing with
both fluctuations and correlations (in some specific scale-free approach)
showing, among other things, that they can make Tsallis entropy additive.
In next Section we show, using simple examples, that fluctuations in
composite (but uncorrelated) systems can induce correlations and $q\neq
1$, i.e., nonextensivity. On the other hand fluctuations in correlated
system can make it apparently uncorrelated and extensive (see also
Appendix A). When applied to multiplicity distributions of particles
produced in collision processes of all kinds it is shown that they
convert the usual Poissonian distribution to the co called Negative
Binomial ones found in all high energy processes \cite{NB} - this is
shown in Section 3. The immediate practical applications of our results
to some recent experimental data are presented in Appendix B.

\section{Tsallis distributions: fluctuations {\it vs} correlations}

\subsection{Two random variable case}

To introduce and discuss correlations one has to deal with at least two
particles. Let $x$ and $y$ denote therefore two independent random
variables, each following its own exponential distribution given by eq.
(\ref{eq:f(x)}) and let their joint probability distribution be given by
$f(x,y) = f(x)\cdot f(y)$. The corresponding Tsallis distribution can be
obtained either by fluctuating the parameter $\lambda$ for each variable
separately (in which case one obtains joint Tsallis distribution for {\it
uncorrelated} random variables) or by fluctuating parameter $\lambda$
jointly for both variables (in which case one gets Tsallis distribution
for {\it correlated} random variables) \cite{BC}. It should be stressed
that fluctuations lead always to Tsallis distribution, irrespectively of
the presence or absence of correlations (their introduction does not
change the statistics). We shall now discuss in more detail the
uncorrelated and correlated cases separately.

\subsubsection{Uncorrelated random variables}

In this case we fluctuate independently parameter $\lambda$ in single
particle probability distributions $f(x)$ and $f(y)$ as given by eq.
(\ref{eq:f(x)}), obtain in this way Tsallis distributions $h(x)$ and
$h(y)$, and finally the joint Tsallis probability distribution in the
form of:
\begin{equation}
h(x,y) = h(x)\cdot h(y) = C^2\cdot\left[ 1 - (1-q)\frac{(x+y)}{\lambda} +
(1-q)^2\frac{xy}{\lambda^2}\right]^{\frac{1}{1-q}}. \label{eq:uncorel}
\end{equation}

However, such distribution of {\it uncorrelated} variables can be also
obtained starting from joint distribution of two {\it correlated}
variables $x$ and $y$ and introducing to it suitable fluctuations in the
way described before. Let us take, for example, the following two
variable distribution,
\begin{equation}
 f(x,y) = \left\{
\frac{\exp[\frac{1}{1-Q}]}{(Q-1)\Gamma\left(0,\frac{1}{Q-1}\right)}\right\}\cdot
\exp\left[ - \frac{x+y}{\lambda'} + (1-Q)\frac{xy}{\lambda\lambda'}
\right] , \label{eq:f(x,y)p}
\end{equation}
where $(Q-1)\ge 0$ describes (negative) correlations between variables
$x$ and $y$ present for $Q \ne 1$. It is characterized by the correlation
coefficient $\rho$:
\begin{eqnarray}
\rho &=& \frac{Cov(x,y)}{\sqrt{Var(x)}\sqrt{Var(y)}} =
  \frac{\langle xy\rangle -
  \langle x\rangle\langle y\rangle}{{\sqrt{Var(x)}\sqrt{Var(y)}}}=
  \label{eq:rho}\\ \nonumber\\ \nonumber\\
&=& \frac{ (Q-1)\exp[\frac{1}{1-Q}] -
E_1\left(\frac{1}{Q-1}\right)\left[1 +
E_1\left(\frac{1}{Q-1}\right)\exp[\frac{1}{1-Q}]\right]}{
(Q-1)\exp[\frac{1}{1-Q}] - QE_1\left(\frac{1}{Q-1}\right)}
\label{eq:rhop}
\end{eqnarray}
\begin{figure}[h]
 \begin{center}
       \epsfig{file=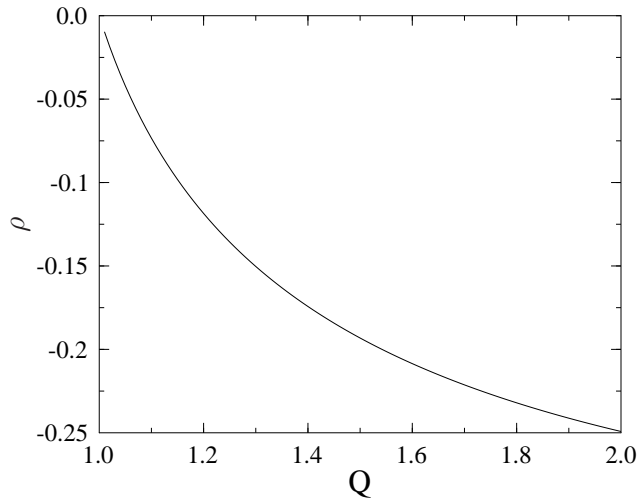, width=100mm}
  \caption{Illustration of dependence of correlation parameter
           $\rho$ as obtained by Eq, (\ref{eq:rhop}) on the
           nonextensivity parameter $Q$.}
 \end{center}
  \label{Figure1}
 \vspace{5mm}
\end{figure}
(we have used here the exponential integral function $E_1(z) =
\int_1^{\infty} \frac{e^{-zt}}{t} dt$ and incomplete gamma function
$\Gamma(0,z) = \int_z^{\infty} \frac{e^{-t}}{t} dt $). Fig. 1 shows how
$\rho$ depends on $Q$. Let us now fluctuate in eq. (\ref{eq:f(x,y)p})
parameter $\lambda'$ in the same way as in eq. (\ref{eq:lambdaprim}) with
$q$ being measure of these fluctuations. It is straightforward to notice
that one introduces in this way {\it positive} correlations between
variables $x$ and $y$ and that for some strength of these positive
correlations given by the condition $q=Q$ they cancel the negative
correlations introduced in eq. (\ref{eq:f(x,y)p}) and we get uncorrelated
variable distribution $h(x,y)$ as given by eq. (\ref{eq:uncorel}). This
example shows that {\it effects of correlations and fluctuations can
cancel each other in final result} \footnote{In Appendix A we provide
another example of this kind when similar approach results also in
extensivity of the final $q$-entropies of the original distributions. It
should be stressed that here $q\ge 1$ whereas in the example presented in
Appendix A we have always $q\le 1$.}.  Let us notice at this point that
correlations assumed in (\ref{eq:f(x,y)p}) which lead to
(\ref{eq:uncorel}) were recently widely used (cf. \cite{BP} and
references therein) as example of the nonextensive rule for addition of
energies resulting in Tsallis distribution. We argue that this is not
true, namely that energies add always additively and formula used in
\cite{BP} ($\kappa(x,y) = x+y + axy$) leads to the formula for joint
probability distribution, as our eqs. (\ref{eq:uncorel}) or
(\ref{eq:f(x,y)p}), which does not provide distribution of the sum of
energies (cf. \cite{BK} and references therein for different mappings
leading to different composition rules).

\subsubsection{Correlated random variables}

To get correlated random variables one fluctuates parameter $\lambda$ in
joint probability distribution of two variables, $\{x,y\}$, as given by
$f(x,y) = f(x)\cdot f(y)$. Performing it in the same way as in
(\ref{eq:h(x)}) one gets the following joint Tsallis probability
distribution:
\begin{eqnarray}
h(x,y) &=& \int_0^{\infty} d\varepsilon \left(
\frac{\varepsilon}{\lambda}\right)^2\exp\left[-
\frac{(x+y)\varepsilon}{\lambda}\right] g(\varepsilon) = \nonumber\\
&=& C_2 \left[1-(1-q)\frac{(x+y)}{\lambda}\right]^{\frac{q}{1-q}}
\qquad{\rm with}\qquad  C_2 = \frac{2-q}{\lambda^2}.
 \label{eq:jointprobdist}
\end{eqnarray}

Notice that marginal probability distributions $h(x) = \int h(x,y) dy  $
and $h(y) = \int h(x,y) dx $ have in this case also form of Tsallis
distributions but with noticeably difference in the exponent, which lacks
now the factor $q$ present in (\ref{eq:jointprobdist}), i.e., they are
identical with eq. (\ref{eq:h(x)}) where exponent is by unity higher than
in (\ref{eq:jointprobdist}) (being equal $1/(1-q) = q/(1-q) + 1$). The
corresponding mean value and variance of the marginal distributions are
equal to
\begin{equation}
\langle x\rangle = \langle y\rangle = \frac{\lambda}{3-2q}\quad {\rm
and}\quad Var(x) = Var(y) = \frac{\lambda^2(2-q)}{(3-2q)^2(4-3q)},
\label{eq:momentsmargxy}
\end{equation}
whereas the corresponding covariance and correlation coefficient
corresponding to this joint probability distribution are equal to
\begin{equation}
Cov(x,y) = \frac{\lambda^2(q-1)}{(3-2q)^2(4-3q)}\quad {\rm and}\quad \rho
= \frac{q-1}{2-q} = \frac{1}{2-q} - 1 .\label{eq:varcorrcoef}
\end{equation}
It means then that correlation coefficient $\rho$ is entirely given by
the parameter $q$, which defines fluctuation of the random variable
$\varepsilon$ (actually, in this case it is just equal to the relative
variance of $\varepsilon$ given by $\omega$, cf. eq. (\ref{eq:defq})).
Notice that we can now express the nonextensivity parameter $q$ via the
correlation coefficient:
\begin{equation}
q = 1 + \frac{\rho}{\rho +1}. \label{eq:qviarho}
\end{equation}
In this way the correlation coefficient defines exponent in the Tsallis
distribution. However, in any possible applications one has to remember
that eq. (\ref{eq:qviarho}) has been obtained only for correlations
caused by fluctuations and therefore cannot be used for estimations of
the role of correlations as such in Tsallis distributions. In Appendix A
we present simple example of correlations making Tsallis entropy additive
(cf. \cite{CT-NEXT2005,CT,TG-MS}).

\subsection{$N$ random variables case}

It is straightforward to proceed to general case of $N$ random variables
$\left\{x_{1,\dots,N}\right\}$ by fluctuating common parameter $\lambda$
in the corresponding initial exponential distribution
$f(\left\{x_{1,\dots,N}\right\}) = \prod_i^N f(x_i)$ (with $f(x_i)$ being
given by eq. (\ref{eq:f(x)})) following prescription given by eq.
(\ref{eq:lambdaprim}). We get in this case following joint probability
distribution:
\begin{eqnarray}
h\left(\{ x_{1,\dots,N} \}\right) &=& \int_0^{\infty} d\varepsilon \left(
\frac{\varepsilon}{\lambda}\right)^N \exp\left[-
\frac{\varepsilon}{\lambda}\cdot \sum_{i=1}^Nx_i \right] g(\varepsilon) =\nonumber\\
&=& C_N\left[1-(1-q)\frac{\sum_{i=1}^N
x_i}{\lambda}\right]^{\frac{1}{1-q}+1 -N} ,\label{eq:hN}
\end{eqnarray}
where
\begin{equation}
C_N=\frac{1}{\lambda^N}\prod_{i=1}^N[(i-2)q - (i-3)] =
\frac{(q-1)^N}{\lambda^N}\cdot \frac{\Gamma\left( N +
\frac{2-q}{q-1}\right)}{\Gamma \left( \frac{2-q}{q-1}\right)}.
\label{eq:CNORMP}
\end{equation}
It is straightforward to check that this distribution leads to marginal
distributions in form of eq. (\ref{eq:h(x)}) for each of variables
considered.

Introducing $N$-particle nonextensivity parameter $q_N$ one can formally
rewrite eq. (\ref{eq:hN}) as:
\begin{equation}
h\left( \{x_{1,\dots,N} \}\right) = C_N\left[ 1\, -\, \frac{1-q_N}{1 +
(N-1)(1-q_N)}\cdot \frac{\sum_{i=1}^N
x_i}{\lambda}\right]^{\frac{1}{1-q_N}} \label{eq:hNA}
\end{equation}
where
\begin{equation}
\frac{1}{1-q_N} - \frac{1}{1-q_1} = 1 - N \quad{\rm or}\quad q_N = 1 +
\frac{q_1-1}{1+ \left(q_1-1\right)(N-1)} \stackrel{\tiny N\rightarrow
\infty}{\Longrightarrow} 1 . \label{eq:qN}
\end{equation}
Notice that, irrespectively of how large are single variable fluctuations
(represented by $q_1$), they disappear in the multi-component systems
with very large number of components $N$ \footnote{~~~~~~~~In
\cite{Parvan}, in thermodynamical context and where $q<1$, one has
$\frac{1}{q_N-1} - \frac{1}{q_1-1} = \frac{3}{2}(1-N)$ instead. Notice
also that if $\left(q_1 - 1 \right)(N-1) >> 1$ one has approximately that
$1/\left(q_N-1\right)  = N\cdot 1/\left(q_1 -1\right)$, which seems to
coincide with the notion of the {\it extensivity} of parameter $\xi =
1/(q-1)$ discussed in \cite{Parvan1,Parvan}. Notice also that
(\ref{eq:hNA}) is multivariable distribution rather then distribution of
the sum of variables discussed recently in the context of the conjectured
$q$-central limit theorem \cite{q-CLT}.}.

\section{Application to multiplicity distributions}

\subsection{Boltzmann distribution and Poisson multiplicity distribution}

Suppose that in some physical process one has $N$ independently produced
secondaries with energies $\{ E_{1,\dots,N}\}$, each distributed
according to Boltzmann distribution, i.e., according to eq.
(\ref{eq:f(x)}) with $x=E_i$ and $\lambda =\langle E_i\rangle$. The
corresponding joint probability distribution is then given by:
\begin{equation}
f\left( \{ E_{1,\dots,N}\} \right) = \frac{1}{\lambda^N}\cdot \exp\left(
- \frac{1}{\lambda}\sum_{i=1}^N E_i\right) = \prod_{i=1}^N \left[
\frac{1}{\lambda}\cdot \exp\left( -
\frac{E_i}{\lambda}\right)\right].\label{eq:jointP}
\end{equation}
For independent $\{ E_{i=1,\dots,N}\}$ the sum $E = \sum_{i=1}^N E_i$ is
then distributed according to gamma distribution,
\begin{equation}
g_N(E) = \frac{1}{\lambda (N-1)!} \cdot
\left(\frac{E}{\lambda}\right)^{N-1}\cdot \exp\left( -
\frac{E}{\lambda}\right)\, =\, g_{N-1}(E)\frac{E}{N-1}, \label{eq:Egamma}
\end{equation}
with distribuant equal to
\begin{equation}
G_N(E) = 1 - \sum_{i=1}^{N-1}\frac{1}{(i-1)!}\cdot
\left(\frac{E}{\lambda}\right)^{i-1}\cdot \exp( - \frac{E}{\lambda}) .
\label{eq:distribuant}
\end{equation}
We look now for such $N$ that $ \sum_{i=0}^N E_i \leq E \le
\sum_{i=0}^{N+1} E_i  $. Their distribution has known Poissonian form
(notice that $E/\lambda = \langle N\rangle$):
\begin{equation}
P(N) = G_{N+1}(E) - G_N(E) =
\frac{\left(\frac{E}{\lambda}\right)^N}{N!}\cdot \exp( - \alpha E) =
\frac{\langle N\rangle ^N}{N!}\cdot \exp( - \langle N\rangle ).
\label{eq:Poisson}
\end{equation}
In other words, whenever we have variables $E_1,E_2,\dots,E_N,E_{N+1},
\dots$ taken from the exponential distribution $f\left(E_i\right)$ and
whenever these variables satisfy the condition $\sum_{i=0}^N E_i \leq E
\le \sum_{i=0}^{N+1} E_i$, then the corresponding multiplicity $N$ has
Poissonian distribution (actually this is precisely the method of
generating Poisson distribution in the numerical Monte-Carlo codes).

\subsection{Tsallis distribution and Negative Binomial multiplicity
distribution}

Suppose now that in another process one has again $N$ particles with
energies $\{ E_{1,\dots,N} \}$ but this time distributed according to
Tsallis distribution as given by eq.(\ref{eq:hN}) (therefore, according
to our previous discussion they cannot be independent but are correlated
in some specific way),
\begin{equation}
h\left( \{E_{1,\dots,N}\}\right) = C_N\left[1-(1-q)\frac{\sum_{i=1}^N
E_i}{\lambda}\right]^{\frac{1}{1-q}+1 -N}, \label{eq:Enq}
\end{equation}
with normalization constant $ C_N$ given by eq. (\ref{eq:CNORMP})). It
means that, according to our reasoning behind eq. (\ref{eq:hN}), there
are some intrinsic (so far unspecified but summarily characterized by the
parameter $q$) fluctuations present in the system under consideration.
Because variables $\{ E_{i=1,\dots,N} \}$ occur in the form of the sum,
$E = \sum_{i=1}^N E_i$, one can perform sequentially integrations of the
joint probability distribution (\ref{eq:Enq}) and, noting that
\begin{equation}
h_N(E) = h_{N-1}(E)\frac{E}{N-1}\qquad{\rm or}\qquad
h_N(E)=\frac{E^{N-1}}{(N-1)!}h\left(\{E_{1,\dots,N}\}\right),
\label{eq:qE123ijkn}
\end{equation}
arrive at formula corresponding to the previous eq. (\ref{eq:Egamma}),
namely
\begin{equation}
h_N(E) = \frac{E^{(N-1)}}{(N-1)!\lambda^N}\prod^N_{i=1}[(i-1)q -
(i-3)]\left[ 1 - (1-q)\frac{E}{\lambda}\right]^{\frac{1}{1-q}+1-N}
\label{eq:Hq} \end{equation}
with distribuant given by
\begin{eqnarray}
\!\!\! H_N(E) &=& 1 -\nonumber\\
     \! \!\!  &-& \sum_{j=1}^{N-1}\left\{
\frac{E^{j-1}}{(j-1)!\lambda^j}\prod_{i=1}^j \left[ (i-1)q -
(i-3)\right]\left[ 1 - (1-q)\frac{E}{\lambda}
\right]^{\frac{1}{1-q}+1-j}\right\} .\label{eq:qdistribuant}
\end{eqnarray}
As before, for energies $E$ satisfying condition $ \sum_{i=0}^N E_i \leq
E \le \sum_{i=0}^{N+1} E_i  $, the corresponding multiplicity
distribution is equal to
\begin{equation}
P(N) = H_{N+1}(E) - H_N(E) \label{eq:qP(N)}
\end{equation}
and is given by the so called Negative Binomial distribution (NBD)
(widely encountered in analyzes of high energy multiparticle production
data of all kinds \cite{NB}):
\begin{eqnarray}
P(N) &=& \frac{(q-1)^N}{N!}\cdot\frac{q-1}{2-q}\cdot
\frac{\Gamma\left(N+1+\frac{2-q}{q-1}\right)}{\Gamma \left(
\frac{2-q}{q-1}\right)}\cdot \left(\frac {E}{\lambda}\right)^N
\left[1-(1-q)\frac{E}{\lambda} \right]^{-N+\frac{1}{1-q}}
\nonumber\\
&=& \frac{\Gamma(N+k)}{\Gamma(N+1)\Gamma(k)}\cdot \frac{\left(
\frac{\langle N\rangle}{k}\right)^N }{\left( 1 + \frac{\langle
N\rangle}{k}\right)^{N+k} } ,\label{eq:NBDfinal}
\end{eqnarray}
where the mean multiplicity and variance are, respectively,
\begin{equation}
\langle N\rangle = \frac{E}{\lambda}; \qquad\quad Var(N) =
\frac{E}{\lambda}\left[ 1 - (1-q)\frac{E}{\lambda}\right] = \langle
N\rangle + \langle N\rangle^2 \cdot (q-1) .\label{eq:meanNVarq}
\end{equation}
(For different way of deriving of NBD by using Tsallis statistics see
\cite{AK}). It is defined by the parameter $k$ equal to:
\begin{equation}
k = \frac{1}{q-1} .\label{eq:k}
\end{equation}
Notice that for $q \rightarrow 1$ one has $k\rightarrow \infty$ and
$P(N)$ becomes Poisson distribution whereas for $q\rightarrow 2$ one has
$k\rightarrow 1$ and we are obtaining geometrical distribution. As we
have notice before, fluctuations described by parameter $q$ result also
in specific correlations described by parameter $\rho$ given by eq.
(\ref{eq:varcorrcoef}). It means that parameter $k$ in NBD can be also
expressed by the correlation coefficient $\rho$ for the two-particle
energy correlations (resulting from intrinsic fluctuations in the
system), namely
\begin{equation}
k = \frac{\rho + 1}{\rho}. \label{eq:krho}
\end{equation}
In should be stressed at this point that result (\ref{eq:k}) coincides
with our previous results in \cite{NBDq} where we have already obtained
NBD from fluctuations of the mean multiplicity in the Poisson
distribution. It means that such fluctuations are equivalent to
fluctuations leading to eq. (\ref{eq:hN}) which, following our reasoning
presented in \cite{WW}, we would like to attribute to the fluctuations of
temperature $T$ for the whole system\footnote{~~~~~~~~ To make this point
more transparent let us notice that because $\langle N\rangle =
E/\lambda$ therefore fluctuation of $\langle N\rangle$ in Poisson
distribution in \cite{NBDq} is equivalent (for fixed $E$ as in our case)
to fluctuation of $1/\lambda$, i.e., in our case to fluctuation of
$\langle E_i\rangle$.}. We would like to close this Section with the
remark that our result harmonizes with the known fact that whereas the
generating function for Poisson distribution is exponential the
corresponding one for the NBD has $q$-exponential form \cite{CaT,AK} (the
relations between generating functions and probability distributions are
the same as between $P(E_i)$ and $P(N)$ in our case, i.e., $P(E_i)$ plays
the role of generating function for distribution $P(N)$).

\section{Summary}

To summarize, let us stress that fluctuations that can be described by
gamma distribution (or equivalent to it in the sense discussed in
\cite{BC} where term of {\it superstatistics} has been coined for this
purpose) lead always to Tsallis distribution. On the other hand, not
every fluctuation results in correlation. This is true only for
fluctuations of the whole multicomponent system $(x+y)$ or
$(\Sigma_ix_i)$. Independent fluctuations of parameters $\lambda_i$ in
$f(x,y) \sim [ \exp( - x/\lambda_x)] \cdot [\exp(-y/\lambda_y)]$ lead to
distribution $h(x,y)=h(x)h(y)$ given by product of two Tsallis
distributions with no correlations between variables $x$ and $y$.
Distributions obtained here differ from some many-particle distributions
for composed systems $\{x_{i=1,\dots,N}\}$ of the form $\exp(\sum_i x_i)
\longrightarrow [1+(1-q)\sum_i x_i]^{\left[ 1/(1-q)\right ]}$, which
occur as apparently natural (and simple) generalization of the
observation that in single component systems fluctuations lead to the
replacement $\exp(x) \longrightarrow [1+(1-q)x]^{\left[1/(1-q)\right]}$,
(and were also used to discuss correlations generated this way \cite{BP})
but they do not lead to correct (marginal) single particle distributions.
Finally, we have proved that energy correlations introduced in
multiparticle system by fluctuations (which can be traced to fluctuations
of temperature in this system, which is the place where energy is
converted into observed particles in process known as {\it
hadronization}) result in changing the corresponding multiplicity
distributions from Poisson to Negative Binomial ones and that this is
equivalent to introducing fluctuations of the mean multiplicities in the
Poisson distribution. The possible application of our findings to some
recent data on multiparticle production processes is presented in
Appendix B.

\appendix
\section{Example of extensive Tsallis entropy}

In \cite{WW} we have shown that nonextensivity leading to Tsallis
statistics \cite{T} and characterized by parameter $q$ can be caused by
some intrinsic fluctuations existing in the physical system under
consideration. The corresponding Tsallis entropy is nonextensive.
However, as discussed in \cite{EuroPhys,CT-NEXT2005,CT,TG-MS} (cf. also
\cite{q-recent}) extensivity depends not only on the specific form of the
entropy function used but also on the composition law according to which
given composed system is formed out of its subsystem, i.e., on their
possible correlations. In fact it is easy to demonstrate
\cite{EuroPhys,CT-NEXT2005,CT,TG-MS,q-recent}) that by introducing to the
physical system some specific correlations (for example, correlations
that are strictly or asymptotically scale invariant
\cite{CT-NEXT2005,CT,TG-MS,q-recent}) one can make the corresponding
entropy becoming extensive. In what follows we shall illustrate this
point by using as example simple gaussian probability distributions for
single and two correlated variables.

Let the single variable $x$ distribution be of the gaussian form (with
$\sigma = \sqrt{Var(x)}$ being a parameter)
\begin{equation}
f(x) = \frac{1}{\sqrt{2\pi}\sigma}\cdot\exp\left( -
\frac{x^2}{2\sigma^2}\right) .\label{eq:ff(x)}
\end{equation}
Let us correlate this variable with another variable, $y$, using two
variable gaussian distribution with correlations provided by parameter
$\rho$ as defined in Eq. (\ref{eq:rho}):
\begin{equation}
f(x,y) = \frac{1}{2\pi \sigma^2\sqrt{1-\rho^2}}\cdot \exp\left[-
\frac{x^2 - 2\rho xy + y^2}{2\sigma^2 \left( 1 - \rho^2\right)}\right].
\label{eq:ff(x,y)}
\end{equation}
It is properly normalized, i.e., $\int \int f(x,y) dx dy =1 $, and has
properly defined marginal probabilities, namely $ \int f(x,y)dy = f(x)$.
It is obvious that $x$ and $y$ are not independent because $ f(x,y) \neq
f(x)f(y) $.

The corresponding Shannon entropies for $f(x)$ and $f(x,y)$ are
\begin{eqnarray}
S_x &=& \frac{1}{2}\left[ 1 + \ln(2\pi ) + 2\ln (\sigma)\right],
\label{eq:Sx}\\
S_{x,y} &=& 1 + \ln (2\pi ) + 2\ln(\sigma) + \ln \sqrt{1-\rho^2},
\label{eq:Sxy}
\end{eqnarray}
respectively, being nonextensive by amount
\begin{equation}
\delta S = S_{x,y} - 2S_x = \ln\sqrt{1-\rho^2} ,\label{eq:deltaS}
\end{equation}
which depends on strength of correlation $\rho$ (notice that $\delta S
\leq 0$) and becoming extensive only for uncorrelated system, i.e., when
$\rho =0$.

\begin{figure}[h]
\noindent
  \begin{minipage}[ht]{70mm}
    \centerline{
       \epsfig{file=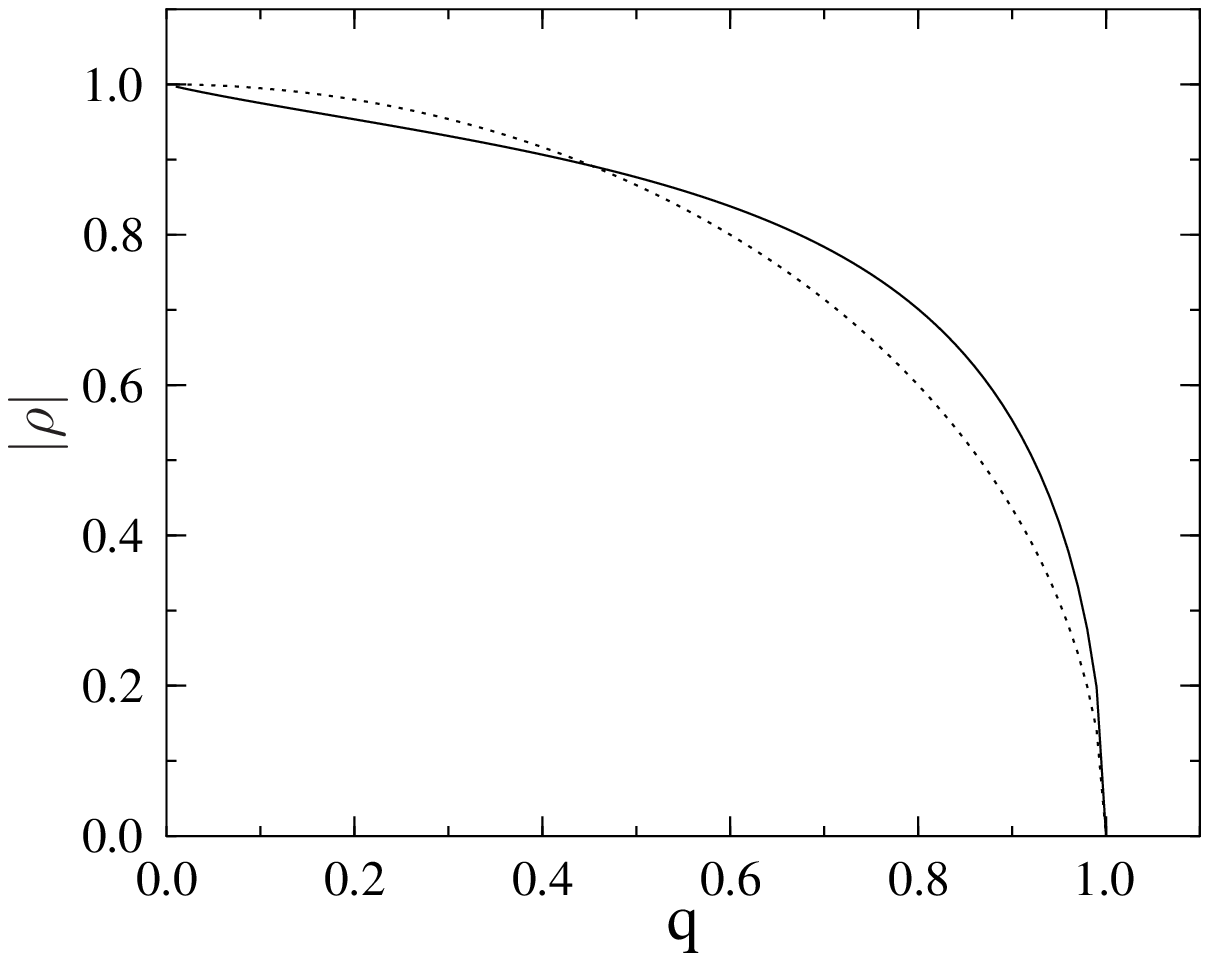, width=70mm}
     }
  \end{minipage}
\hfill
  \begin{minipage}[ht]{70mm}
    \centerline{
       \epsfig{file=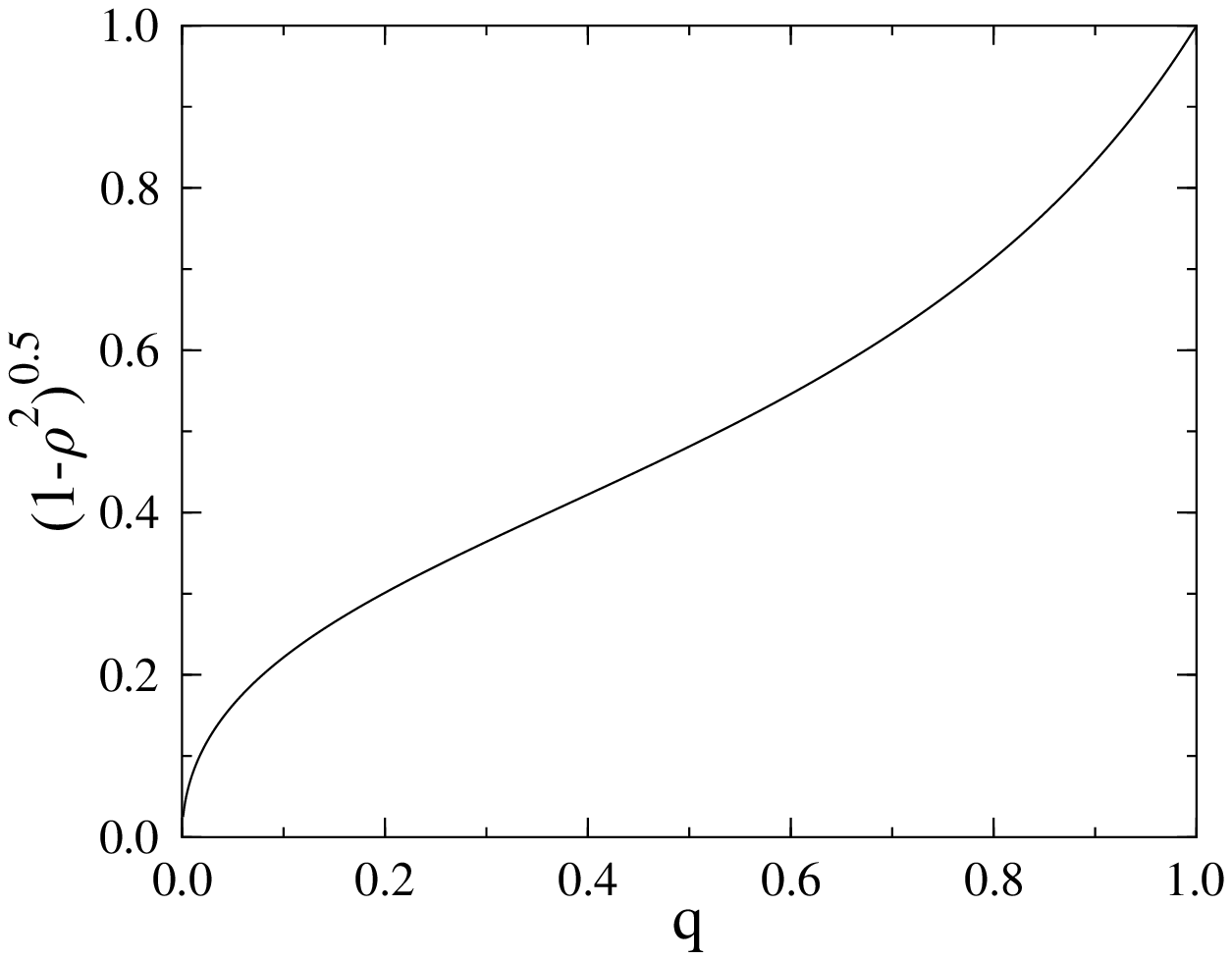, width=70mm}
     }
  \end{minipage}
  \caption{Left panel: $|\rho|$ versus $q$: full line - exact formula
           (\ref{eq:qrho}); dotted line - approximate formula
           (\ref{eq:approxqrho}).
           Right panel: $\sqrt{1-\rho^2}$ versus $q$. Notice that $0 \leq\, q\, \leq 1
           $.
  }
  \label{Fig2}
  \vspace{5mm}
\end{figure}

The corresponding Tsallis entropies are equal to:
\begin{eqnarray}
T_x &=& \frac{1}{q-1}\left[ 1 - \frac{(2\pi
)^{\frac{1-q}{2}}\sigma^{1-q}}{\sqrt{q}}\right]  \label{eq:Tx}\\
T_{x,y} &=& \frac{1}{q-1} \left[ 1 - \frac{(2\pi )^{1-q}\left(\ \sigma^2
\sqrt{1-\rho^2}\right)^{1-q}}{q}\right]   \label{eq:Txy}\\
\delta T &=& T_{x,y} - 2 T_x =\nonumber\\
&=& \frac{1}{q(1-q)} \left[ q + (2\pi )^{1-q}\left( \sigma^2
\sqrt{1-\rho^2}\right)^{1-q} -
2^{\frac{3-q}{2}}\pi^{\frac{1-q}{2}}\sigma^{1-q}\sqrt{q}  \right].
\label{eq:deltaT}
\end{eqnarray}
Notice that for uncorrelated variables, i.e., for $\rho=0$, one gets in
this case the usual result for Tsallis entropy:
\begin{equation}
\delta T = (1-q) T_x^2 .\label{eq:zerpcorT}
\end{equation}
However, it is obvious from eq. (\ref{eq:deltaT}) (cf. also Fig.
\ref{Fig2}) that one can always find such value of correlation $\rho$ for
which $\delta T =0$ and Tsallis entropy becomes extensive:
\begin{equation}
\sqrt{1-\rho^2} = \left[ 2(2\pi)^{\frac{q-1}{2}} \sqrt{q} - (2\pi)^{q-1}q
\right]^{\frac{1}{1-q}},\label{eq:qrho}
\end{equation}
or, in crude approximation,
\begin{equation}
\sqrt{1-\rho^2} \approx q; \qquad 0 \leq\, q\, \leq 1 .
\label{eq:approxqrho}
\end{equation}
Notice that, according to eq. (\ref{eq:deltaS}), both positive and
negative correlations result in some (negative) imbalance of Shannon and
Tsallis entropy.
One has to stress at this point that what we have shown here is only a
kind of formal exercise presented for illustration use only. It does not
prove that correlations of some type lead to Tsallis type distributions.
Variables can be correlated in any way and this does not depend on the
distribution.

\section{Example of possible practical applications}

Results presented in this paper have practical application in the field
of high energy multiparticle production reactions, especially those
originated by collisions of heavy nuclei, which are of particular
interest as potential source of production of new state of matter, the so
called {\it Quark Gluon Plasma} (QGP) (cf. references in
\cite{B,J,A,Ad}). Recently number of works \cite{B,J,A,Ad} have
demonstrated the existence in such reactions event-by-event fluctuations
of the average transverse momenta $\langle p\rangle$ per event. The
following quantities were considered: $Var\left(\langle p\rangle
\right)/\langle \langle p\rangle \rangle^2$ and $\langle \Delta p_i
\Delta p_j\rangle/\langle \langle p\rangle \rangle^2$. These quantities,
as we advocate, are fully determined by $\omega$ as defined by eq.
(\ref{eq:defq}),which in our case translates into fluctuations of the
temperature $T$ of hadronizing system - a vital observable when searching
for QGP\footnote{~~~~~~~~Generally speaking, analysis of transverse
momenta $p_T$ alone indicates very small fluctuations of $T$. On the
other hand, as reported in \cite{MR}, the measured fluctuations of
multiplicities of produced secondaries are large (i.e., multiplicity
distributions are substantially broader than Poissonian). Our analysis of
NBD applied to observed multiplicity distributions show that this can
result in large fluctuations of $T$, cf. \cite{NUWW}.}. The results
obtained in \cite{B,J,A,Ad} can be interpreted as fluctuations of the
temperature $T$ of the hadronic matter being produced (these are
fluctuations for the whole event or its part (cluster) but not for the
particular (single) particles in an event).

To show this let us consider the case of $N_{ev}$ events with $N_k$
particles in the $k^{\rm th}$ event. Introducing the following notation:
\begin{eqnarray}
C_k &=& \sum_i^{N_k}\, \sum_j^{N_k}\, \left(p_i - \langle \langle
p\rangle\rangle \right)\cdot \left(p_j - \langle \langle p\rangle\rangle
\right), \label{eq:Ck}\\
\langle \langle p \rangle \rangle &=& = \frac{1}{N_{ev}}\,
\sum_k^{N_{ev}}\,\langle p\rangle_k;\qquad {\rm where}\qquad \langle
p\rangle_k = \frac{1}{N_k}\, \sum_i^{N_k}\,p_i , \label{eq:llprr}
\end{eqnarray}
we have that
\begin{eqnarray}
C &=& \langle \Delta p_i\Delta p_j\rangle = \frac{1}{N_{ev}}\,
\sum_k^{N_{ev}}\, \frac{C_k}{N_k\left(N_k - 1\right)} \label{eq:C11}.
\end{eqnarray}
Adding and subtracting the same term $\langle p\rangle_k^2$ it can be
written as
\begin{eqnarray}
 C &=& \frac{1}{N_{ev}}\, \sum_k^{N_{ev}}\left[ \frac{1}{N_k\left(N_k
-1\right)}\sum_i^{N_k}\sum_j^{N_k}p_ip_j + \left(\langle p\rangle^2_k -
\langle p\rangle_k^2\right) - \langle\langle p\rangle\rangle^2\right].
\label{eq:C22}
\end{eqnarray}
If particles in the event are independent then
\begin{equation}
\frac{1}{N_k\left(N_k -1\right)}\sum_i^{N_k}\sum_j^{N_k}p_ip_j  - \langle
p\rangle_k^2 = 0 \label{eq:C333}
\end{equation}
and we have that
\begin{equation}
C = \frac{1}{N_{ev}} \sum^{N_{ev}}_k\left( \langle p\rangle^2_k - \langle
\langle p\rangle\rangle^2 \right ) = Var\left(\langle p\rangle \right),
\label{eq:C444}
\end{equation}
or that
\begin{equation}
\frac{C}{\langle \langle p\rangle\rangle^2} = \frac{Var(\langle
p\rangle)}{\langle \langle p \rangle\rangle^2} = \frac{Var(T)}{\langle
T\rangle^2} = \omega. \label{eq:C555}
\end{equation}
The above formulas can be checked against data obtained at Relativistic
Heavy Ion Collider (RHIC) at Brookhaven Nat. Lab where $Au$ nuclei are
impinging at each other with center of mass energy $200$ GeV per nucleon.
Data taken by STAR experiment \cite{A} for centrality $30-40$\%
($N_{part} \sim 100$) give $\omega=4\cdot 10^{-4}$. The respective values
for data obtained by other experiment, PHENIX, \cite{Ad} are: $2.2\cdot
10^{-4}$, $2.4\cdot 10^{-4}$, $3.6\cdot 10^{-4}$ and $4.9\cdot 10^{-4}$
for the respective centralities: $0-5$\%, $0-10$\%, $10-20$\% and
$20-30$\%. The more detailed analysis of the RHIC data along the line
presented here is, however, out of the scope of the present paper and
will be presented elsewhere.\\


Partial support of the Polish Ministry of Science and Higher Education
(grants 1P03B12730 (ZW) and 621/E-78/SPB/CERN/P-03/DWM 52/2004-2006  and
1 P03B022 30 (GW)) is acknowledged.\\

\end{document}